\documentclass[twocolumn,showpacs,preprintnumbers,amsmath,amssymb,epsfig]{revtex4-2}

\usepackage{graphicx}
\usepackage{dcolumn}
\usepackage{bm}
\usepackage{epsfig}

\usepackage{float}
\usepackage{amsmath}
\usepackage{epsfig,floatflt}
\usepackage{subfigure}
\usepackage[usenames]{color}
\usepackage{rotating}
\usepackage{comment} 
\usepackage{hyperref}

\newcommand{\mii}{Mg\,\textsc{ii}}
\newcommand{\civ}{C\,\textsc{iv}}

\usepackage[T1]{fontenc}
\usepackage{scalerel}
\usepackage{tikz}
\usetikzlibrary{svg.path}

\definecolor{orcidlogocol}{HTML}{A6CE39}
\tikzset{
  orcidlogo/.pic={
    \fill[orcidlogocol] svg{M256,128c0,70.7-57.3,128-128,128C57.3,256,0,198.7,0,128C0,57.3,57.3,0,128,0C198.7,0,256,57.3,256,128z};
    \fill[white] svg{M86.3,186.2H70.9V79.1h15.4v48.4V186.2z}
                 svg{M108.9,79.1h41.6c39.6,0,57,28.3,57,53.6c0,27.5-21.5,53.6-56.8,53.6h-41.8V79.1z M124.3,172.4h24.5c34.9,0,42.9-26.5,42.9-39.7c0-21.5-13.7-39.7-43.7-39.7h-23.7V172.4z}
                 svg{M88.7,56.8c0,5.5-4.5,10.1-10.1,10.1c-5.6,0-10.1-4.6-10.1-10.1c0-5.6,4.5-10.1,10.1-10.1C84.2,46.7,88.7,51.3,88.7,56.8z};
  }
}

\newcommand\orcidicon[1]{\href{https://orcid.org/#1}{\mbox{\scalerel*{
\begin{tikzpicture}[yscale=-1,transform shape]
\pic{orcidlogo};
\end{tikzpicture}
}{|}}}}

\begin{document}


\title{Using non-DESI data to confirm and strengthen the DESI 2024 spatially-flat $w_0w_a$CDM cosmological parameterization result}

\author{Chan-Gyung Park$^{\orcidicon{0000-0002-3076-2781}}$}%
 \email{park.chan.gyung@gmail.com}
\affiliation{Division of Science Education and Institute of Fusion Science, Jeonbuk National University, Jeonju 54896, Republic of Korea}%

\author{Javier de Cruz P\'erez$^{\orcidicon{0000-0001-8603-5447}}$}%
 \email{jadecruz@ucm.es}
\affiliation{Departamento de F\'isica Te\'orica, Universidad Complutense de Madrid, 28040, Madrid, Spain}%

\author{Bharat Ratra$^{\orcidicon{0000-0002-7307-0726}}$}%
 \email{ratra@phys.ksu.edu}
\affiliation{Department of Physics, Kansas State University, 116 Cardwell Hall, Manhattan, KS 66506, USA}%

\date{\today}

\begin{abstract}

We use a combination of Planck cosmic microwave background (CMB) anisotropy data and non-CMB data that include Pantheon+ type Ia supernovae (SNIa), Hubble parameter [$H(z)$], growth factor ($f\sigma_8$) measurements, and a collection of baryon acoustic oscillation (BAO) data, but not recent DESI 2024 BAO measurements, to confirm the DESI 2024 (DESI+CMB+PantheonPlus) data compilation support for dynamical dark energy with an evolving equation of state parameter $w(z) = w_0 + w_a z/(1+z)$. From our joint compilation of CMB and non-CMB data, in a spatially-flat cosmological model, we obtain $w_0 = -0.850 \pm 0.059$ and $w_a = -0.59^{+0.26}_{-0.22}$ and find that this dynamical dark energy is favored over a cosmological constant by $\sim 2\sigma$. Our data constraints on the flat $w_0w_a$CDM parameterization are slightly more restrictive than the DESI 2024 constraints, with the DESI 2024 and our values of $w_0$ and $w_a$ differing by $-0.27\sigma$ and $0.44\sigma$, respectively. Our data compilation slightly more strongly favors the flat $w_0w_a$CDM model over the flat $\Lambda$CDM model than does the DESI 2024 data compilation. We note that our CMB and non-CMB data $w_0w_a$CDM parameterization cosmological constraints are discrepant at 2.7$\sigma$, a little larger than the 1.9$\sigma$ discrepancy between DESI DR1 BAO and CMB data flat $\Lambda$CDM model cosmological constraints. We also show that if we remove the Pantheon+ SNIa contribution from the non-CMB data, for the $w_0w_a$CDM parameterization we still find tension between P18 and non-CMB data (2.5$\sigma$) and P18+lensing and non-CMB data (2.4$\sigma$). Even after the exclusion of Pantheon+ SNIa data the $\Lambda$CDM model is still disfavoured at $\sim 2\sigma$ c.l. 

\end{abstract}
\pacs{98.80.-k, 95.36.+x}

\maketitle

\maxdeadcycles=200

\section{Introduction}
\label{sec:Introduction} 

In the six-parameter spatially-flat $\Lambda$CDM cosmological model, \cite{Peebles:1984ge}, the observed currently accelerated cosmological expansion is a general-relativistic gravitational effect sourced by the cosmological constant $\Lambda$ that currently dominates the cosmological energy budget with pressure-less cold dark matter (CDM) being the next largest contributor to the current energy budget. While the currently-standard spatially-flat $\Lambda$CDM model is generally consistent with most observational constraints, some current measurements might be incompatible with the predictions of this model, \cite{Perivolaropoulos:2021jda, Moresco:2022phi, Abdalla:2022yfr, Hu:2023jqc}.

Recent DESI baryon acoustic oscillation (BAO) measurements, \cite{DESI:2024mwx}, might be incompatible with the predictions of the standard flat $\Lambda$CDM model. In an analysis of the spatially-flat $w_0w_a$CDM cosmological parameterization where dynamical dark energy is modelled as a fluid with an evolving equation of state parameter
$w(z) = w_0 + w_az/(1+z)$, \cite{Chevallier:2000qy, Linder:2002et}, DESI+CMB+PantheonPlus data (described below) favor $w_0 = -0.827 \pm 0.063$ and $w_a = -0.75^{+0.29}_{-0.25}$, approximately $2\sigma$ away from the cosmological constant point at $w_0 = -1 $ and $w_a = 0$. (We note that in the following we use $w_0$ and $w$ interchangeably.) For other discussions of constraints on dynamical dark energy see \cite{Sola:2016hnq, Ooba:2017lng, Ooba:2018dzf, Park:2018fxx,SolaPeracaula:2018wwm, Park:2019emi, Cao:2020jgu, Khadka:2020hvb, Cao:2022ugh, Dong:2023jtk, VanRaamsdonk:2023ion, deCruzPerez:2024abc} and references therein. 

Here we use the P18+lensing+non-CMB data set of \cite{deCruzPerez:2024abc} (see Sec.\ II for the meaning and composition of the P18, lensing, and non-CMB data sets), where non-CMB data includes Pantheon+ SNIa data, and in particular BAO data that does not include the recent DESI 2024 BAO measurements. We confirm the DESI 2024 support for dynamical dark energy with an evolving equation of state parameter $w(z) = w_0 + w_az/(1+z)$, but now with $w_0 = -0.850 \pm 0.059$ and $w_a = -0.59^{+0.26}_{-0.22}$, which suggests that dynamical dark energy is preferred over a cosmological constant by $\sim 2\sigma$. Our P18+lensing+non-CMB data constraints on the flat $w_0w_a$CDM model are slightly more restrictive than those derived in \cite{DESI:2024mwx}, while also slightly more strongly favoring the flat $w_0w_a$CDM model over the flat $\Lambda$CDM model than found in \cite{DESI:2024mwx}. 

While interesting, these results are not statistically significant. More importantly $w(z) = w_0 + w_az/(1+z)$ is a parameterization and not a physically consistent dynamical dark energy model. The simplest physically-consistent dynamical dark energy models use a dynamical scalar field $\phi$ with a self-interaction potential energy density $V(\phi)$ as dynamical dark energy, \cite{Peebles:1987ek, Ratra:1987rm}. For recent discussions of dynamical scalar field dark energy models in the context of DESI 2024 measurements, see \cite{Tada:2024znt, Yin:2024hba, Berghaus:2024kra}. For other discussions of the DESI 2024 results, see \cite{Wang:2024hks, Luongo:2024fww, Cortes:2024lgw, Colgain:2024xqj, Wang:2024rjd, Wang:2024dka}.

In Sec.\ \ref{sec:Data} we provide brief details of the data sets we use to constrain cosmological parameters in, and test the performance of, the flat $w_0w_a$CDM model. In Sec.\ \ref{sec:Methods} we briefly summarize the main features of the flat $w_0w_a$CDM model and the analysis techniques we use. Our results are presented and discussed in Sec.\ \ref{sec:ResultsandDiscussion}, and we conclude in Sec.\ \ref{sec:Conclusion}.  

\section{Data}
\label{sec:Data}

In this work CMB and non-CMB data sets are used to constrain the parameters of a dynamical dark energy model. The data sets we use in our analyses here are described in detail in Sec.\ II of \cite{deCruzPerez:2024abc} and outlined in what follows. We account for all known data covariances. 

For the CMB data, we use the Planck 2018 TT,TE,EE+lowE (P18) CMB temperature and polarization power spectra alone as well as jointly with the Planck lensing potential (lensing) power spectrum \cite{Planck:2018nkj,Planck:2018vyg}. 

The non-CMB data set we use is the non-CMB (new) data compilation of \cite{deCruzPerez:2024abc}, which is comprised of 

\begin{itemize}

\item 16 BAO data points, spanning $0.122 \le z \le 2.334$ which include results from the 6dFGS+SDSS MGS \cite{Carter:2018vce}, BOSS Galaxy \cite{Gil-Marin:2020bct}, eBOSS LRG \cite{Gil-Marin:2020bct,Bautista:2020ahg}, DES Y3 \cite{DES:2021wwk}, eBOSS Quasar \cite{Hou:2020rse,Neveux:2020voa}, and Ly$\alpha$-forest \cite{duMasdesBourboux:2020pck} surveys. These data are listed in Table I of \cite{deCruzPerez:2024abc}. We do not use DESI 2024 BAO data, \cite{DESI:2024mwx}. 

\item A 1590 SNIa data point subset of the Pantheon+ compilation \cite{Brout:2022vxf}, retaining only SNIa with $z > 0.01$ to mitigate peculiar velocity correction effects. These data span $0.01016 \le z \le 2.26137$,

\item 32 Hubble parameter [$H(z)$] measurements, obtained using the differential age technique, spanning $0.070 \le z \le 1.965$. These data points are listed in Table 1 of \cite{Cao:2023eja} and in Table II of \cite{deCruzPerez:2024abc} and the original references are \cite{Zhang:2012mp,Moresco:2020fbm,Simon:2004tf,Ratsimbazafy:2017vga,Stern:2009ep,Borghi:2021rft,Moresco:2022phi}. 

\item An additional nine (non-BAO) growth rate ($f\sigma_8$) data points, 
spanning $0.013 \le z \le 1.36$ from the ALFALFA \cite{Avila:2021dqv}, IRAS \cite{Turnbull:2011ty,Hudson:2012gt}, 6dFGS+SDSS \cite{Said:2020epb}, SDSS DR7 \cite{Shi:2017qpr}, eBOSS LRG \cite{Simpson:2015yfa,Blake:2013nif}, VIPERS \cite{Mohammad:2018mdy}, and FastSound \cite{Okumura:2015lvp} surveys that are listed in Table III of \cite{deCruzPerez:2024abc}.

\end{itemize}

The difference between our non-CMB data compilation and the data used in the DESI 2024 analysis \cite{DESI:2024mwx} is that our compilation uses a different BAO dataset compared to the DESI 2024 analysis, though both use the same Pantheon+ SNIa sample (although we also consider a non-CMB data compilation that excludes the Pantheon+ SNIa). Additionally, our non-CMB data includes growth rate and $H(z)$ measurements.

We use five individual and combined data sets to constrain the flat $w_0w_a$CDM and flat $\Lambda$CDM models: P18 data, P18+lensing data, non-CMB data, P18+non-CMB data, and P18+lensing+non-CMB data.

\section{Methods}
\label{sec:Methods}

The methods we use are described in Sec.\ III of  \cite{deCruzPerez:2024abc}. A brief summary follows.

To determine quantitatively how tightly these observational data constrain the cosmological model parameters, we use the \texttt{CAMB}/\texttt{COSMOMC} program (October 2018 version) \cite{Challinor:1998xk,Lewis:1999bs,Lewis:2002ah}. \texttt{CAMB} computes the evolution of model spatial inhomogeneities and makes theoretical predictions which depend on cosmological parameters while \texttt{COSMOMC} compares these predictions to observational data, using the Markov chain Monte Carlo (MCMC) method, to determine cosmological parameter likelihoods. The MCMC chains are assumed to have converged when the Gelman and Rubin $R$ statistic satisfies $R-1 < 0.01$. For each model and data set, we use the converged MCMC chains, with the \texttt{GetDist} code \cite{Lewis:2019xzd}, to compute the average values, confidence intervals, and likelihood distributions of model parameters. 

In the flat $\Lambda$CDM model, the six primary cosmological parameters are conventionally chosen to be the current value of the physical baryonic matter density parameter $\Omega_b h^2$, the current value of the physical cold dark matter density parameter $\Omega_c h^2$, the angular size of the sound horizon at recombination $100\theta_{\text{MC}}$, the reionization optical depth $\tau$, the primordial scalar-type perturbation power spectral index $n_s$, and the power spectrum amplitude $\ln(10^{10}A_s)$, where $h$ is the Hubble constant in units of 100 km s$^{-1}$ Mpc$^{-1}$. We assume flat priors for these parameters, non-zero over: $0.005 \le \Omega_b h^2 \le 0.1$, $0.001 \le \Omega_c h^2 \le 0.99$, $0.5 \le 100\theta_\textrm{MC} \le 10$, $0.01 \le \tau \le 0.8$, $0.8 \le n_s \le 1.2$, and $1.61 \le \ln(10^{10} A_s) \le 3.91$. In the $w_0 w_a$CDM model dynamical dark energy is assumed to be a fluid with an evolving equation of state parameter (fluid pressure to energy density ratio) $w(z) = w_0 + w_az/(1+z)$, \cite{Chevallier:2000qy, Linder:2002et}, and for the additional dark energy equation of state parameters we also adopt flat priors non-zero over $-3.0 \le w_0 \le 0.2$ and $-3 < w_a < 2$. We adopt the parametrized post-Friedmann dark energy model \cite{Fang:2008sn} to allow for a dark energy fluid whose equation of state parameter crosses $w=-1$. When we estimate parameters using non-CMB data, we fix the values of $\tau$ and $n_s$ to those obtained from P18 data (since these parameters cannot be determined solely from non-CMB data) and constrain the other cosmological parameters. Additionally, we also present constraints on three derived parameters, namely the Hubble constant $H_0$, the current value of the non-relativistic matter density parameter $\Omega_m$, and the amplitude of matter fluctuations $\sigma_8$, which are obtained from the primary parameters of the cosmological model. In all cases, we restrict the range of the derived parameter $H_0$ to $0.2 \le h \le 1$.


\begin{table*}
	\caption{Mean and 68\% (or 95\% indicated between parentheses when the value is provided) confidence limits of flat $w_0 w_a$CDM model parameters
        from non-CMB, P18, P18+lensing, P18+non-CMB, and P18+lensing+non-CMB data.
        $H_0$ has units of km s$^{-1}$ Mpc$^{-1}$. We also include the values of $\chi^2_{\text{min}}$, DIC, and AIC and the differences with respect to the $\Lambda$CDM model, denoted by $\Delta\chi^2_{\text{min}}$, $\Delta$DIC, and $\Delta$AIC, respectively.}
\begin{ruledtabular}
\begin{tabular}{lccccc}
  Parameter                     &  Non-CMB                     & P18                         &  P18+lensing               &  P18+non-CMB            & P18+lensing+non-CMB     \\[+1mm]
 \hline \\[-1mm]
  $\Omega_b h^2$                & $0.0315 \pm 0.0043$          & $0.02240 \pm 0.00015$       & $0.02243 \pm 0.00015$      &  $0.02245 \pm 0.00014$  &  $0.02244 \pm 0.00014$  \\[+1mm]
  $\Omega_c h^2$                & $0.0990^{+0.0061}_{-0.011}$  & $0.1199 \pm 0.0014$         & $0.1192 \pm 0.0012$        &  $0.1190 \pm 0.0011$    &  $0.1191 \pm 0.0010$  \\[+1mm]
  $100\theta_\textrm{MC}$       & $1.0218^{+0.0087}_{-0.011}$  & $1.04094 \pm 0.00031$       & $1.04101 \pm 0.00031$      &  $1.04101 \pm 0.00030$  &  $1.04100 \pm 0.00029$  \\[+1mm]
  $\tau$                        & $0.0540$                     & $0.0540 \pm 0.0079$         & $0.0523 \pm 0.0074$        &  $0.0529 \pm 0.0077$    &  $0.0534 \pm 0.0072$    \\[+1mm]
  $n_s$                         & $0.9654$                     & $0.9654 \pm 0.0043$         & $0.9669 \pm 0.0041$        &  $0.9672 \pm 0.0040$    &  $0.9670 \pm 0.0039$    \\[+1mm]
  $\ln(10^{10} A_s)$            & $3.60\pm 0.24$ ($>3.13$)     & $3.043 \pm 0.016$           & $3.038 \pm 0.014$          &  $3.039 \pm 0.016$      &  $3.040 \pm 0.014$      \\[+1mm]
  $w_0$                         & $-0.876 \pm 0.055$           & $-1.25^{+0.43}_{-0.56}$     & $-1.24^{+0.44}_{-0.56}$    &  $-0.853 \pm 0.061$     &  $-0.850 \pm 0.059$     \\[+1mm]  
  $w_a$                         & $0.10^{+0.32}_{-0.20}$       & $-1.3 \pm 1.2$ ($< 1.13$)   & $-1.2 \pm 1.3$ ($<1.19$)   &  $-0.57^{+0.27}_{-0.23}$&  $-0.59^{+0.26}_{-0.22}$     \\[+1mm]  
 \hline \\[-1mm]
  $H_0$                         & $69.8\pm 2.4$                & $84 \pm 11$ ($> 64.5$)      & $84 \pm 11$ ($>64.7$)      &  $67.81 \pm 0.64$       &  $67.80 \pm 0.64$       \\[+1mm]
  $\Omega_m$                    & $0.2692^{+0.0086}_{-0.015}$  & $0.213^{+0.016}_{-0.070}$   & $0.213^{+0.017}_{-0.071}$  &  $0.3092 \pm 0.0063$    &  $0.3094 \pm 0.0063$    \\[+1mm]
  $\sigma_8$                    & $0.823^{+0.031}_{-0.027}$    & $0.955^{+0.11}_{-0.050}$    & $0.945^{+0.11}_{-0.048}$   &  $0.810 \pm 0.011$      &  $0.8108 \pm 0.0091$    \\[+1mm]
      \hline\\[-1mm]
  $\chi_{\textrm{min}}^2$       & $1457.16$                    & $2761.18$                   & $2770.39$                  & $4234.18$               &  $4243.01$              \\[+1mm]
  $\Delta\chi_{\textrm{min}}^2$ & $-12.77$                     & $-4.62$                     & $-4.32$                    & $-6.06$                 &  $-6.25$              \\[+1mm]
  $\textrm{DIC}$                & $1470.93$                    & $2815.45$                   & $2824.19$                  & $4290.48$               &  $4298.75$              \\[+1mm]
  $\Delta\textrm{DIC}$          & $-7.18$                      & $-2.48$                     & $-2.26$                    & $-1.85$                 &  $-2.45$                \\[+1mm]
  $\textrm{AIC}$                & $1469.16$                    & $2819.18$                   & $2828.39$                  & $4292.18$               &  $4301.01$              \\[+1mm]
  $\Delta\textrm{AIC}$          & $-8.77$                      & $-0.62$                     & $-0.32$                    & $-2.05$                 &  $-2.25$                \\[+1mm]
\end{tabular}
\\[+1mm]
\begin{flushleft}
\end{flushleft}
\end{ruledtabular}
\label{tab:results_flat_w0waCDM}
\end{table*}


For the flat tilted (with $n_s$ not necessarily unity) $w_0 w_a$CDM model the primordial scalar-type energy density perturbation power spectrum is
\begin{equation}
    P_\delta (k) = A_s \left( \frac{k}{k_0} \right)^{n_s},
\label{eq:powden-flat}
\end{equation}
where $k$ is the wavenumber and $n_s$ and $A_s$ are the spectral index and the amplitude of the spectrum at pivot scale $k_0=0.05~\textrm{Mpc}^{-1}$. This power spectrum is generated by quantum fluctuations during an early epoch of power-law inflation in a spatially-flat inflation model powered by a scalar field inflaton potential energy density that is an exponential function of the inflaton \cite{Lucchin:1984yf, Ratra:1989uv, Ratra:1989uz}.

To quantify how relatively well each model fits the data set under study, we use the differences in the Akaike information criterion ($\Delta$AIC) and the deviance information criterion ($\Delta$DIC) between the information criterion (IC) values for the flat $w_0 w_a$CDM model and the flat $\Lambda$CDM model. See Sec.\ III of \cite{deCruzPerez:2024abc} for a fuller discussion. According to the Jeffreys' scale (which can vary slightly depending on the reference considered), when $-2 \leq \Delta\textrm{IC}<0$ there is {\it weak} evidence in favor of the model under study, while when $-6 \leq \Delta\textrm{IC} < -2$ there is {\it positive} evidence, when $-10\leq\Delta\textrm{IC} < -6$ there is {\it strong} evidence, and when $\Delta\textrm{IC} < -10$ there is {\it very strong} evidence in favor of the model under study relative to the flat $\Lambda$CDM model. This scale also holds when $\Delta\textrm{IC}$ is positive, but then the tilted flat $\Lambda$CDM model is favored over the model under study.

To quantitatively compare how consistent the cosmological parameter constraints (for the same model) derived from two different data sets are, we use two estimators. The first is the DIC based $\log_{10}\mathcal{I}$, see \cite{Joudaki:2016mvz} and Sec.\ III of \cite{deCruzPerez:2024abc}. When the two data sets are consistent $\log_{10}\mathcal{I}>0$ while $\log_{10}\mathcal{I}<0$ means that the two data sets are inconsistent. According to the Jeffreys' scale the degree of consistency or inconsistency between two data sets is {\it substantial} if $\lvert \log_{10}\mathcal{I} \rvert >0.5$, {\it strong} if $\lvert \log_{10}\mathcal{I} \rvert >1$, and {\it decisive} if $\lvert \log_{10}\mathcal{I} \rvert >2$, \cite{Joudaki:2016mvz}. The second estimator is the tension probability $p$ and the related, Gaussian approximation, "sigma value" $\sigma$, see \cite{Handley:2019pqx, Handley:2019wlz, Handley:2019tkm} and Sec.\ III of \cite{deCruzPerez:2024abc}. $p=0.05$  and $p=0.003$ correspond to 2$\sigma$ and 3$\sigma$ Gaussian standard deviation. 

\begin{figure*}[htbp]
\centering
\mbox{\includegraphics[width=170mm]{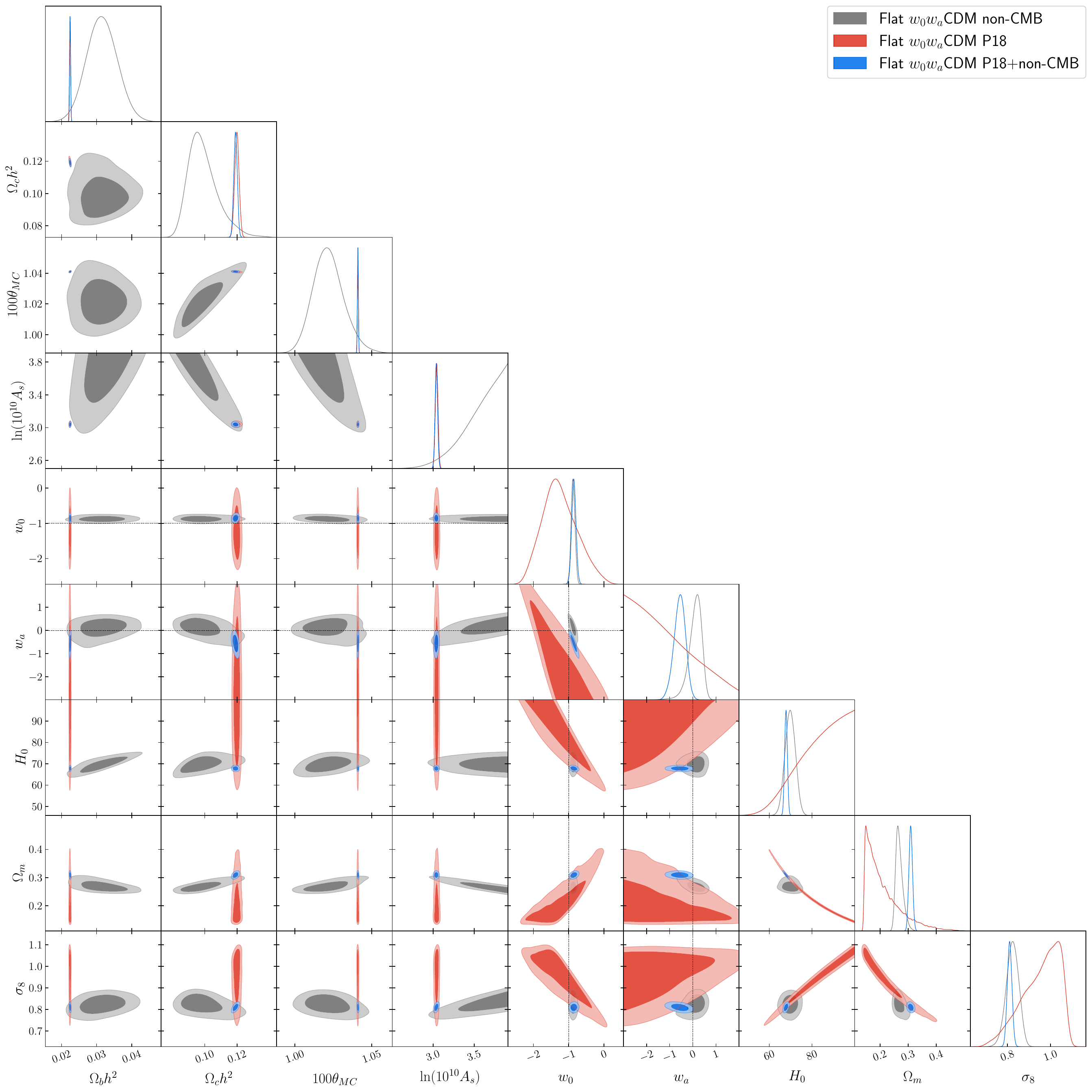}}
        \caption{One-dimensional likelihoods and 1$\sigma$ and $2\sigma$ likelihood confidence contours of flat $w_0 w_a$CDM model parameters favored by non-CMB, P18, and P18+non-CMB data sets. We do not show $\tau$ and $n_s$, which are fixed in the non-CMB data analysis.
}
\label{fig:flat_w0waCDM_P18_nonCMB23v2}
\end{figure*}

\begin{figure*}[htbp]
\centering
\mbox{\includegraphics[width=170mm]{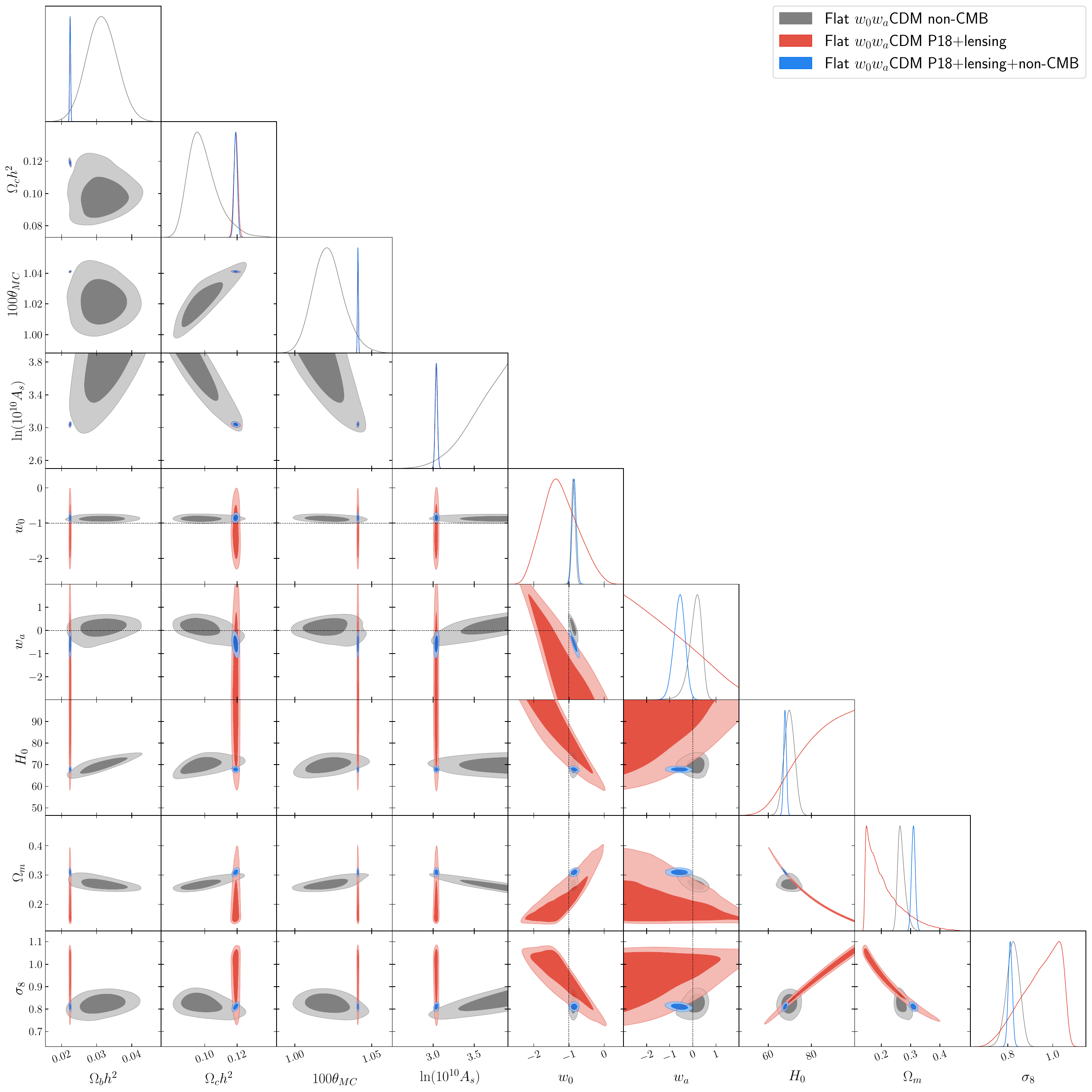}}
\caption{One-dimensional likelihoods and 1$\sigma$ and $2\sigma$ likelihood confidence contours of flat $w_0 w_a$CDM model parameters favored by non-CMB, P18+lensing, P18+lensing+non-CMB data sets. We do not show $\tau$ and $n_s$, which are fixed in the non-CMB data analysis.
}
\label{fig:flat_w0waCDM_P18_lensing_nonCMB23v2}
\end{figure*}


\section{Results and Discussion}
\label{sec:ResultsandDiscussion}

\begin{figure*}[htbp]
\centering
\mbox{\includegraphics[width=85mm]{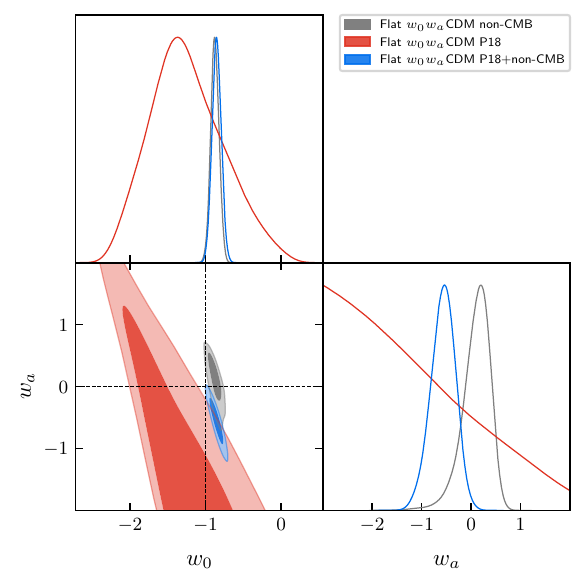}}
\mbox{\includegraphics[width=85mm]{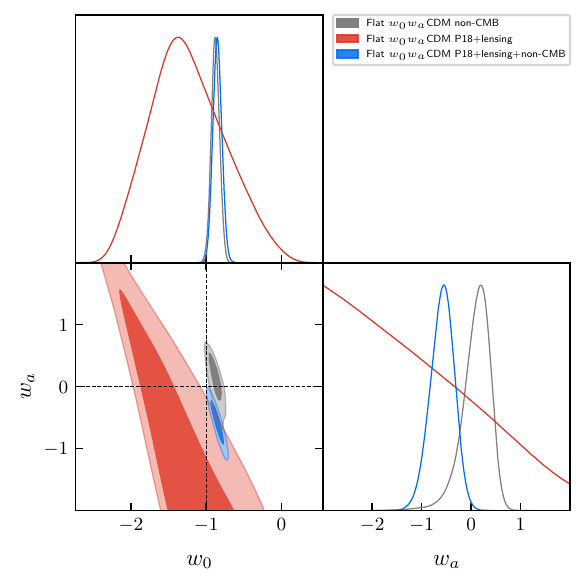}}
        \caption{One-dimensional likelihoods and 1$\sigma$ and $2\sigma$ likelihood confidence contours of $w_0$ and $w_a$ parameters in the flat $w_0 w_a$CDM model favored by (left) non-CMB, P18, and P18+non-CMB data sets, and (right) non-CMB, P18+lensing, and P18+lensing+non-CMB data sets.
}
\label{fig:flat_w0waCDM_P18_nonCMB23v2_w0wa}
\end{figure*}

\begin{figure*}[htbp]
\centering
\mbox{\includegraphics[width=170mm]{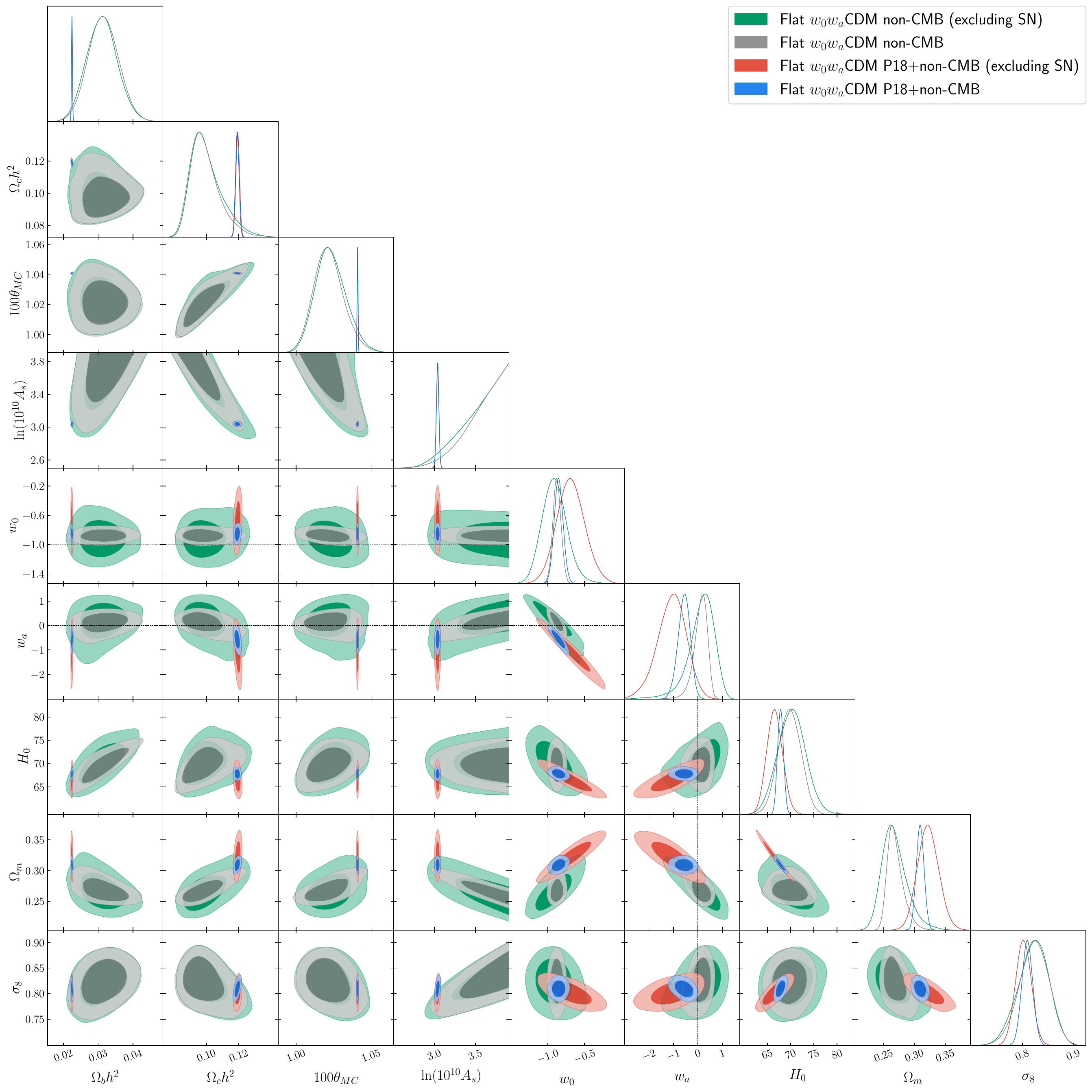}}
        \caption{One-dimensional likelihoods and 1$\sigma$ and $2\sigma$ likelihood confidence contours of flat $w_0 w_a$CDM model parameters favored by non-CMB (excluding SN), non-CMB, P18+non-CMB (excluding SN), and P18+non-CMB data sets. We do not show $\tau$ and $n_s$, which are fixed in the non-CMB data analysis.
}
\label{fig:flat_w0waCDM_P18_nonCMB23v2_NoSN}
\end{figure*}

\begin{figure*}[htbp]
\centering
\mbox{\includegraphics[width=170mm]{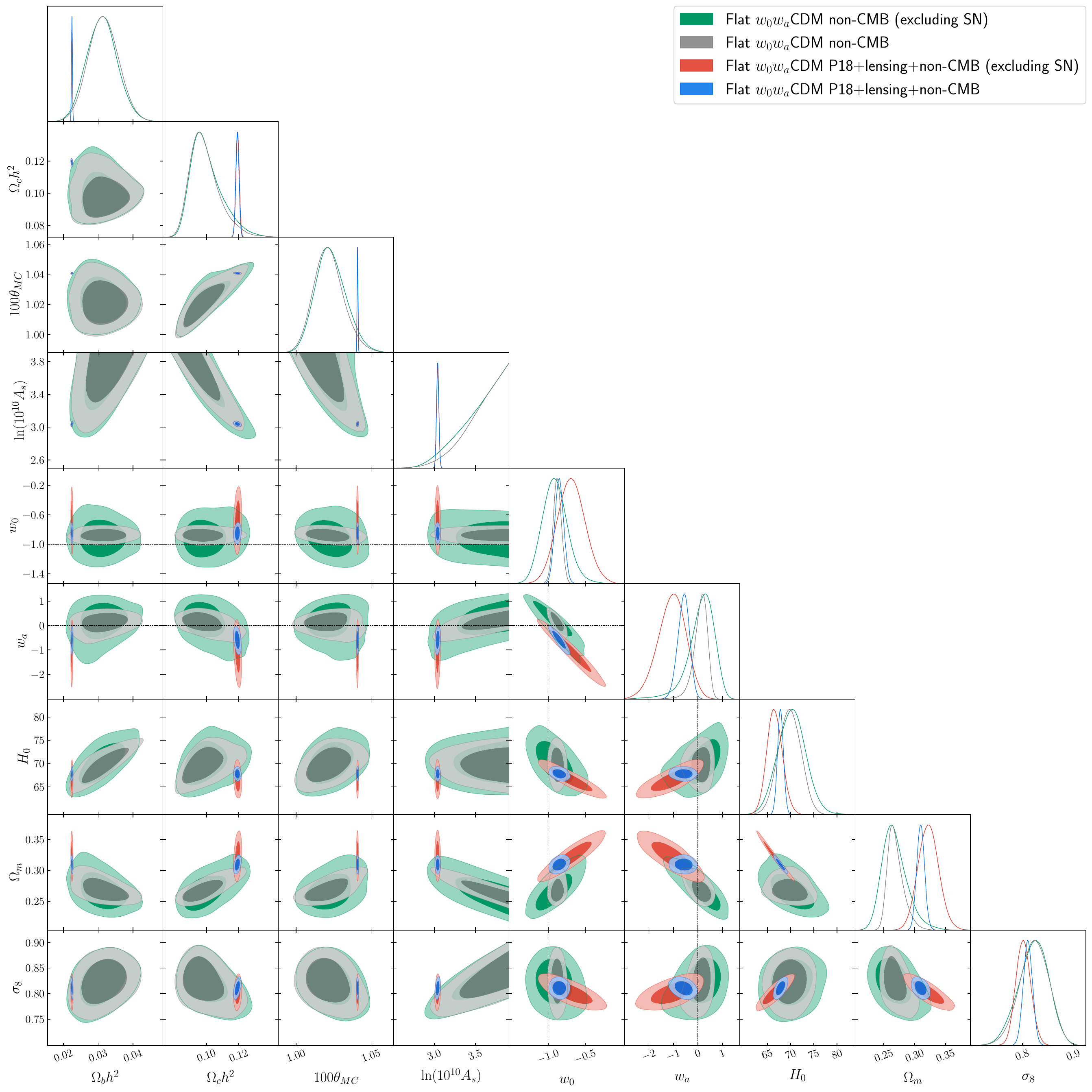}}
\caption{One-dimensional likelihoods and 1$\sigma$ and $2\sigma$ likelihood confidence contours of flat $w_0 w_a$CDM model parameters favored by non-CMB (excluding SN), non-CMB, P18+lensing+non-CMB (excluding SN), and P18+lensing+non-CMB data sets. We do not show $\tau$ and $n_s$, which are fixed in the non-CMB data analysis.
}
\label{fig:flat_w0waCDM_P18_lensing_nonCMB23v2_NoSN}
\end{figure*}

\begin{figure*}[htbp]
\centering
\mbox{\includegraphics[width=85mm]{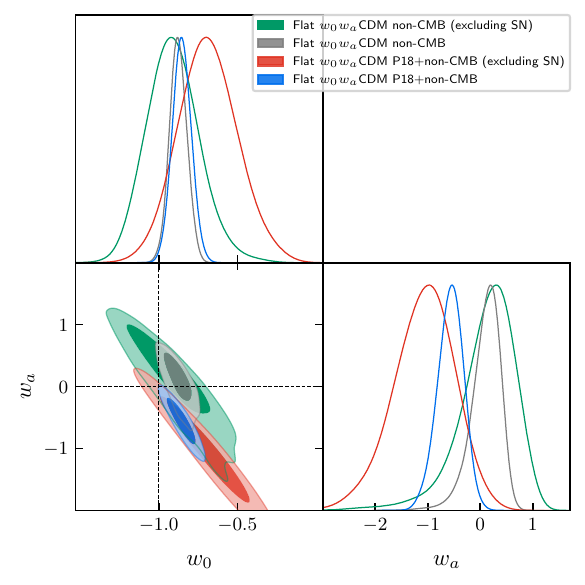}}
\mbox{\includegraphics[width=85mm]{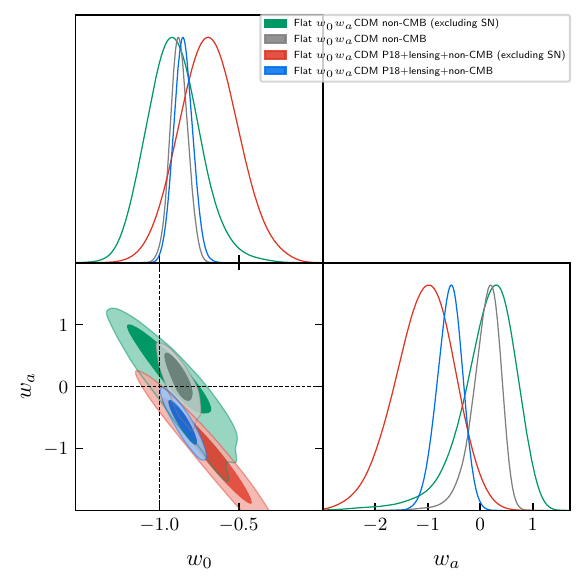}}
        \caption{One-dimensional likelihoods and 1$\sigma$ and $2\sigma$ likelihood confidence contours of $w_0$ and $w_a$ parameters in the flat $w_0 w_a$CDM model favored by (left) non-CMB (excluding SN), non-CMB, and P18+non-CMB (excluding SN), and P18+non-CMB data sets, and (right) non-CMB (excluding SN), non-CMB, P18+lensing+non-CMB (excluding SN), P18+lensing+non-CMB data sets.
}
\label{fig:flat_w0waCDM_P18_nonCMB23v2_w0wa_NoSN}
\end{figure*}

Cosmological parameter constraints are listed in Tables \ref{tab:results_flat_w0waCDM} and \ref{tab:results_flat_w0waCDM_NoSN} and shown in Figs.\ \ref{fig:flat_w0waCDM_P18_nonCMB23v2}, \ref{fig:flat_w0waCDM_P18_lensing_nonCMB23v2}, \ref{fig:flat_w0waCDM_P18_nonCMB23v2_NoSN}, and \ref{fig:flat_w0waCDM_P18_lensing_nonCMB23v2_NoSN} with just the $w_0-w_a$ panels of these figures reproduced in Figs.\ \ref{fig:flat_w0waCDM_P18_nonCMB23v2_w0wa} and \ref{fig:flat_w0waCDM_P18_nonCMB23v2_w0wa_NoSN}. Values of the statistical estimators used to assess consistency between P18 and non-CMB data cosmological constraint results and between P18+lensing and non-CMB data results are listed in Tables \ref {tab:consistency_w0waCDM} and \ref{tab:consistency_w0waCDM_NoSN}, while $\Delta \chi^2_{\rm min}$, $\Delta$AIC, and $\Delta$DIC values are listed in Tables \ref{tab:results_flat_w0waCDM} and \ref{tab:results_flat_w0waCDM_NoSN}. 

From Table \ref{tab:results_flat_w0waCDM} and Figs.\ \ref{fig:flat_w0waCDM_P18_nonCMB23v2} and \ref{fig:flat_w0waCDM_P18_lensing_nonCMB23v2} we see that non-CMB data provide significantly more restrictive constraints on $w_0$ and $w_a$, as well as on derived parameters $H_0$, $\Omega_m$, and $\sigma_8$, than do P18 or P18+lensing data. This is very similar to what happens in the XCDM (or $w$CDM) model, to which the $w_0w_a$CDM model studied here reduces when $w_a = 0$, see the discussion in Sec.\ IV.B of \cite{deCruzPerez:2024abc}.

\begin{table}
\caption{Consistency check parameter $\log_{10} \mathcal{I}$ and tension parameters $\sigma$ and $p$ for P18 vs.\ non-CMB data sets and P18+lensing vs.\ non-CMB data sets in the flat $w_0 w_a$CDM model.
}
\begin{ruledtabular}
\begin{tabular}{lcc}
   Data                          &  P18 vs non-CMB  & P18+lensing vs non-CMB  \\[+1mm]
 \hline \\[-1mm]
  $\log_{10} \mathcal{I}$        &   $-0.891$       &  $-0.787$       \\[+1mm]
  $\sigma$                       &   $2.801$        &  $2.653$        \\[+1mm]
  $p$ (\%)                       &   $0.509$        &  $0.798$        \\[+1mm]
\end{tabular}
\\[+1mm]
\end{ruledtabular}
\label{tab:consistency_w0waCDM}
\end{table}


\begin{table*}
	\caption{Mean and 68\% (or 95\% indicated between parentheses when the value is provided) confidence limits of flat $w_0 w_a$CDM model parameters
        from non-CMB (excluding SN), P18+non-CMB (excluding SN), and P18+lensing+non-CMB (excluding SN) data.
        $H_0$ has units of km s$^{-1}$ Mpc$^{-1}$. We also include the values of $\chi^2_{\text{min}}$, DIC, and AIC and the differences with respect to the $\Lambda$CDM model, denoted by $\Delta\chi^2_{\text{min}}$, $\Delta$DIC, and $\Delta$AIC, respectively.}
\begin{ruledtabular}
\begin{tabular}{lccc}
  Parameter                     &  Non-CMB (excluding SN)          &  P18+non-CMB (excluding SN)            & P18+lensing+non-CMB (excluding SN)     \\[+1mm]
 \hline \\[-1mm]
  $\Omega_b h^2$                & $0.0311 \pm 0.0044$              &  $0.02243 \pm 0.00014$   &  $0.02243 \pm 0.00014$  \\[+1mm]
  $\Omega_c h^2$                & $0.1001^{+0.0061}_{-0.012}$      &  $0.1194 \pm 0.0011$     &  $0.1193 \pm 0.0010$  \\[+1mm]
  $100\theta_\textrm{MC}$       & $1.0232^{+0.0091}_{-0.011}$      &  $1.04098 \pm 0.00030$   &  $1.04097 \pm 0.00030$  \\[+1mm]
  $\tau$                        & $0.0540$                         &  $0.0525 \pm 0.0077$     &  $0.0526 \pm 0.0074$    \\[+1mm]
  $n_s$                         & $0.9654$                         &  $0.9666 \pm 0.0040$     &  $0.9665 \pm 0.0038$    \\[+1mm]
  $\ln(10^{10} A_s)$            & $3.56\pm 0.26$ ($>3.06$)         &  $3.039 \pm 0.016$       &  $3.039 \pm 0.014$      \\[+1mm]
  $w_0$                         & $-0.91^{+0.15}_{-0.18}$          &  $-0.69 \pm 0.19$        &  $-0.69 \pm 0.19$     \\[+1mm]  
  $w_a$                         & $0.13^{+0.61}_{-0.36}$           &  $-1.07^{+0.62}_{-0.53}$ &  $-1.07^{+0.61}_{-0.52}$     \\[+1mm]  
 \hline \\[-1mm]
  $H_0$                         & $70.2 \pm 3.0$                   &  $66.5 \pm 1.7$          &  $66.5 \pm 1.7$       \\[+1mm]
  $\Omega_m$                    & $0.268^{+0.015}_{-0.023}$        &  $0.322 \pm 0.017$       &  $0.322 \pm 0.017$    \\[+1mm]
  $\sigma_8$                    & $0.821^{+0.033}_{-0.027}$        &  $0.802 \pm 0.015$       &  $0.802 \pm 0.015$    \\[+1mm]
      \hline\\[-1mm]
  $\chi_{\textrm{min}}^2$       & $45.24$                          & $2821.43$                &  $2829.99$              \\[+1mm]
  $\Delta\chi_{\textrm{min}}^2$ & $-5.65$                          & $-4.19$                  &  $-4.92$              \\[+1mm]
  $\textrm{DIC}$                & $59.36$                          & $2877.49$                &  $2886.03$              \\[+1mm]
  $\Delta\textrm{DIC}$          & $+0.81$                          & $-0.00$                  &  $-0.64$                \\[+1mm]
  $\textrm{AIC}$                & $57.24$                          & $2879.43$                &  $2887.99$              \\[+1mm]
  $\Delta\textrm{AIC}$          & $-1.65$                          & $-0.19$                  &  $-0.92$                \\[+1mm]
\end{tabular}
\\[+1mm]
\begin{flushleft}
\end{flushleft}
\end{ruledtabular}
\label{tab:results_flat_w0waCDM_NoSN}
\end{table*}


\begin{table}
\caption{Consistency check parameter $\log_{10} \mathcal{I}$ and tension parameters $\sigma$ and $p$ for P18 vs.\ non-CMB (excluding SN) data sets and P18+lensing vs.\ non-CMB (excluding SN) data sets in the flat $w_0 w_a$CDM model.
}
\begin{ruledtabular}
\begin{tabular}{lcc}
   Data                          &  P18 vs non-CMB  & P18+lensing vs non-CMB \\[+1mm]
                                 &  (excluding SN)  & (excluding SN) \\[+1mm]

 \hline \\[-1mm]
  $\log_{10} \mathcal{I}$        &   $-0.582$       &  $+0.289$       \\[+1mm]
  $\sigma$                       &   $2.507$        &  $2.448$        \\[+1mm]
  $p$ (\%)                       &   $1.220$        &  $1.439$        \\[+1mm]
\end{tabular}
\\[+1mm]
\end{ruledtabular}
\label{tab:consistency_w0waCDM_NoSN}
\end{table}


Table \ref {tab:consistency_w0waCDM} shows that non-CMB and P18+lensing data constraints are incompatible at $2.7\sigma$ in the flat $w_0w_a$CDM model (with non-CMB and P18 data being slightly more incompatible at $2.8\sigma$) according to the second ($p$ and $\sigma$) estimator we use; according to $\log_{10}\mathcal{I}$ there is a substantial tension between the two data sets which indicates that the results must be interpreted with caution. This should be compared to the $1.2\sigma$ compatibility and $3.6\sigma$ incompatibility between these two data sets in the flat $\Lambda$CDM model and the flat XCDM model (the dark energy parameterization with a constant $w_0$ equation of state parameter that can differ from $-1$), respectively, see Tables X and XIV of \cite{deCruzPerez:2024abc}, where according to $\log_{10}\mathcal{I}$ there is substantial consistency (flat $\Lambda$CDM) and decisive inconsistency (flat XCDM) between these data sets. One may conclude that these data rule out the flat $w_0w_a$CDM model at $2.7\sigma$ (or $2.8\sigma)$, but given the current state of the field we instead conclude that P18 or P18+lensing data and non-CMB data are compatible at better than $3\sigma$ in the flat $w_0w_a$CDM model and so can be jointly used to constrain cosmological parameters in this model. In the following discussion we will focus more on the P18+lensing+non-CMB data results, as that is the largest data set we use. 

In order to check whether Pantheon+ SNIa data are driving the aforementioned tensions between P18 and non-CMB and P18+lensing and non-CMB we remove from the analysis its contribution, see Tables \ref{tab:results_flat_w0waCDM_NoSN} and \ref{tab:consistency_w0waCDM_NoSN} and Figs.\ \ref{fig:flat_w0waCDM_P18_nonCMB23v2_NoSN}, \ref{fig:flat_w0waCDM_P18_lensing_nonCMB23v2_NoSN} and \ref{fig:flat_w0waCDM_P18_nonCMB23v2_w0wa_NoSN}. In this case we get for P18 vs.\ non-CMB (excluding SNIa) a $\sigma=2.507$ tension between the two data sets whereas for the case P18+lensing vs.\ non-CMB (excluding SNIa) the discrepancy is $\sigma=2.448$. In light of these results we conclude that the SNIa data are not responsible for the tension between the high-redshift and the low-redshift data. On the other hand, the evidence found in favor of the $w_0w_a$CDM parameterization fitting these data (relative to the flat $\Lambda$CDM model fit) subsides when the SNIa contribution is excluded from the analysis. For the three different data set combinations that include non-CMB (excluding SNIa) data, when compared to the $\Lambda$CDM performance, we get, non-CMB (excluding SNIa) ($\Delta\text{DIC}=+0.81$), P18+non-CMB (excluding SNIa) ($\Delta\text{DIC}= 0.00$), and P18+lensing+non-CMB (excluding SNIa) ($\Delta\text{DIC}=-0.64$). The impact of the exclusion of the SNIa data is also reflected in a shift in the values of the equation of state parameters ($w_0,w_a$). In particular, while for the non-CMB, P18+non-CMB, and P18+lensing+non-CMB data sets we get ($w_0=-0.876\pm 0.055$, $w_a=0.10^{+0.32}_{-0.20}$), ($w_0=-0.853\pm 0.061$, $w_a=-0.57^{+0.27}_{-0.23}$), and ($w_0=-0.850\pm 0.059$, $w_a=-0.59^{+0.26}_{-0.22}$), for the corresponding data sets excluding Pantheon+ SNIa data we obtain ($w_0=-0.91^{+0.15}_{-0.18}$, $w_a=0.13^{+0.61}_{-0.36}$), ($w_0=-0.69\pm 0.19$, $w_a=-1.07^{+0.62}_{-0.53}$), and ($w_0=-0.69\pm 0.19$, $w_a=-1.07^{+0.61}_{-0.52}$) with the differences between the corresponding pair of values being ($0.21\sigma$, $-0.06\sigma$), ($-0.82\sigma$, $0.76\sigma$), and ($-0.80\sigma$, $0.74\sigma$). Importantly, from Figs.\ \ref{fig:flat_w0waCDM_P18_nonCMB23v2_w0wa_NoSN}, we see that even when Pantheon+ SNIa data are excluded from the analyses, the flat $\Lambda$CDM model with $w_0 = -1$ and $w_a = 0$ is still $\sim2\sigma$ away from the best-fit models for the red P18+non-CMB (excluding SNIa) data and the red P18+lensing+non-CMB (excluding SNIa) data contours. 

In an attempt to better support our choice to somewhat downplay the $2.7\sigma$ and $2.8\sigma$ incompatibilities discussed in the previous paragraph, we note that in addition to the flat $\Lambda$CDM model issues alluded to in Sec.\ \ref{sec:Introduction} there are two less widely discussed puzzles with some of the data sets we use. One has to do with P18 data in the seven-parameter flat $\Lambda$CDM+$A_L$ model where the phenomenological lensing consistency parameter $A_L$ is introduced to rescale the amplitude of the gravitational potential power spectrum, \cite{Calabreseetal2008}. Here $A_L=1$ corresponds to the theoretically predicted (using the best-fit cosmological parameter values) amount of weak lensing of the CMB anisotropy. When analyzing P18 data one discovers that $A_L>1$ is favored over $A_L =1$ at $2.8\sigma$, \cite{Calabreseetal2008, Planck:2018vyg, deCruzPerez:2022hfr, deCruzPerez:2024abc}. We however note that a recent analysis of updated PR4 Planck data, \cite{Tristram:2023haj}, finds $A_L > 1$ is favored over $A_L = 1$ by only $0.75\sigma$. Another issue is that some SNIa data tend to favor higher values of $\Omega_m$ than do other data. For example, in the flat $\Lambda$CDM model P18+lensing data give $\Omega_m  = 0.3153 \pm 0.0073$, \cite{Planck:2018vyg}, while Pantheon+ SNIa data give $\Omega_m  = 0.332 \pm 0.020$, \cite{Cao:2023eja}, which are not that discrepant, but other SNIa data (which we do not use here) give higher values, $\Omega_m  = 0.356^{+0.028}_ {-0.026}$, \cite{Rubin:2023ovl}, and  $\Omega_m  = 0.352 \pm 0.017$, \cite{DES:2024tys}. So it is probably not inconceivable that there might be a few as yet undiscovered systematics in some cosmological data.   

Comparing the flat $\Lambda$CDM model cosmological parameter values constrained by the P18+lensing+non-CMB (new) data, listed in the right column of the upper panel of Table IV of \cite{deCruzPerez:2024abc}, to those for the same data but for the flat $w_0w_a$CDM model shown in the right column of Table \ref{tab:results_flat_w0waCDM} here, we find that the six common primary parameter values are in good agreement, with the differences being $0.26\sigma$ for $\Omega_b h^2$, $-0.47\sigma$ for $\Omega_c h^2$, $0.22\sigma$ for $100\theta_{\text{MC}}$, $0.35\sigma$ for $\tau$, $0.28\sigma$ for $n_s$, and $0.30\sigma$ for $\ln(10^{10}A_s)$, with equally small derived-parameter differences of $0.34\sigma$ for $H_0$, $-0.44\sigma$ for $\Omega_m$, and $-0.29\sigma$ for $\sigma_8$. It is encouraging that current data compilations are able to provide almost cosmological-model-independent main cosmological parameter constraints.

From the P18+lensing+non-CMB data set in the flat $w_0w_a$CDM model we get $H_0=67.80\pm 0.64$ km s$^{-1}$ Mpc$^{-1}$, which agrees with the median statistics result $H_0=68\pm 2.8$ km km s$^{-1}$ Mpc$^{-1}$ \cite{Chen:2011ab,Gottetal2001,Calabreseetal2012}, as well as with some other local measurements including the flat $\Lambda$CDM model value of \cite{Cao:2023eja} $H_0=69.5\pm 2.4$ km s$^{-1}$ Mpc$^{-1}$ from a joint analysis of $H(z)$, BAO, Pantheon+ SNIa, quasar angular size, reverberation-measured \mii\ and \civ\ quasar, and 118 Amati correlation gamma-ray burst data, and the local $H_0=69.8\pm 1.7$ km s$^{-1}$ Mpc$^{-1}$ from TRGB and SNIa data \cite{Freedman:2021ahq}, but is in tension with the local $H_0=73.04\pm 1.04$ km s$^{-1}$ Mpc$^{-1}$ measured using Cepheids and SNIa data \cite{Riess:2021jrx}, also see \cite{Chen:2024gnu}. And the flat $w_0w_a$CDM model P18+lensing+non-CMB data value $\Omega_m = 0.3094\pm 0.0063$ also agrees well with the flat $\Lambda$CDM model value of $\Omega_m = 0.313\pm 0.012$ of \cite{Cao:2023eja} (for the data set listed above used to determine $H_0$).

The DESI collaboration, \cite{DESI:2024mwx}, compile the DESI+CMB+PantheonPlus data set, from DESI 2024 BAO measurements (DESI), P18 power spectra measurements \cite{Planck:2018nkj, Planck:2018vyg} combined with updated Planck and Atacama Cosmology Telescope lensing potential power spectrum measurements \cite{ACT:2023kun, ACT:2023dou, ACT:2023ubw} (CMB), and Pantheon+ SNIa \cite{Brout:2022vxf} (PantheonPlus). From the DESI+CMB+PantheonPlus data in the flat $w_0w_a$CDM model they measure (and list in their Table 3) $w_0 = -0.827 \pm 0.063$, $w_a = -0.75^{+0.29}_{-0.25}$, $H_0=68.03\pm 0.72$ km s$^{-1}$ Mpc$^{-1}$, and  $\Omega_m = 0.3085\pm 0.0068$, differing by $-0.27\sigma$, $0.44\sigma$, $-0.24\sigma$, and $0.097\sigma$, respectively from our somewhat more restrictive P18+lensing+non-CMB values of $w_0 = -0.850 \pm 0.059$, $w_a = -0.59^{+0.26}_{-0.22}$, $H_0=67.80\pm 0.64$ km s$^{-1}$ Mpc$^{-1}$, and $\Omega_m = 0.3094\pm 0.0063$ listed in the right column of our Table \ref{tab:results_flat_w0waCDM}. It is important to remember that the data sets employed in \cite{DESI:2024mwx} and here are not identical, since here we also include data from cosmic chronometers and a set of $f\sigma_8$ data points computed at different redshifts as well as other BAO data points that are not used in \cite{DESI:2024mwx}. However, both data sets share P18 CMB data. 

Comparing our $w_0$--$w_a$ likelihood contours of the flat $w_0 w_a$CDM model for the P18+lensing+non-CMB data, shown in the right panel of Fig.\ \ref{fig:flat_w0waCDM_P18_nonCMB23v2_w0wa}, to the corresponding DESI+CMB+PantheonPlus blue contours in the right panel of Fig.\ 6 of \cite{DESI:2024mwx}, we see that the upper left vertex of our $2\sigma$ blue contour almost touches the flat $\Lambda$CDM model point of $w_0 = -1$ and $w_a = 0$, while the corresponding DESI+CMB+PantheonPlus point is slightly removed from the flat $\Lambda$CDM point towards slightly more negative values of $w_0$ and $w_a$. The major axis of our $2\sigma$ contour is roughly half as long as the corresponding DESI+CMB+PantheonPlus one, reflecting the greater constraining power of our data compilation. Figure \ref{fig:flat_w0waCDM_P18_nonCMB23v2} indicates how the P18 likelihood contours clearly favor the phantom region for the equation of state parameters of dark energy, as well as regions of high values of $H_0$ and low values of $\Omega_m$. This has to do with the existing degeneracy between these parameters that cannot be broken using the CMB data alone. The inclusion of non-CMB data results in a shrinkage of the favored cosmological parameters regions which drives $H_0$ to lower values and $\Omega_m$ to higher values and decreases the evidence in favor of phantom dark energy.

From Table \ref{tab:results_flat_w0waCDM} we see that for P18+lensing+non-CMB data the flat $w_0w_a$CDM model is positively favored over the flat $\Lambda$CDM model by $\Delta {\rm DIC} = -2.45$, slightly more favored by these data than by the DESI+CMB+PantheonPlus data compilation, where $\Delta {\rm DIC} = -2.0$, \cite{DESI:2024mwx}, is on the borderline and indicates weak evidence for the flat $w_0w_a$CDM model.

\section{Conclusion}
\label{sec:Conclusion}

Using the P18+lensing+non-CMB data set of \cite{deCruzPerez:2024abc}, that is about as independent of DESI 2024 data \cite{DESI:2024mwx} as reasonably possible (there is some spatial overlap at lower-$z$ between some of the BAO data sets), we have confirmed the DESI 2024 finding that a dynamical dark energy density fluid parameterized by an evolving equation of state parameter $w(z) = w_0 + w_az/(1+z)$ with $w_0 = -0.850 \pm 0.059$ and $w_a = -0.59^{+0.26}_{-0.22}$, is favored over a cosmological constant by $\sim 2\sigma$. Our P18+lensing+non-CMB data constraints on the flat $w_0w_a$CDM model cosmological parameters are slightly more restrictive than those derived from DESI+CMB+PantheonPlus data of \cite{DESI:2024mwx}. P18+lensing+non-CMB data also slightly more strongly favor the flat $w_0w_a$CDM model over the flat $\Lambda$CDM model than do DESI+CMB+PantheonPlus data. Within the context of the $w_0w_a$CDM parameterization we find a tension of 2.7$\sigma$ between CMB+lensing and non-CMB data cosmological constraints. As an additional test we have removed the Pantheon+ SNIa data from the analysis in order to determine whether this data set is causing these tensions. It turns out that the tension between P18 and non-CMB (excluding SNIa) cosmological data constraints still remain at a similar level as the cases when SNIa data are included and the $\Lambda$CDM model is still disfavored at $\sim 2\sigma$ c.l. Also, we have recently shown \cite{Chan-GyungPark:2024brx} that when the lensing consistency parameter $A_L$ is allowed to vary, the $w_0w_a$CDM$+A_L$ parameterization, for P18+lensing+non-CMB data, favors $A_L > 1$ at $\sim 2\sigma$ and still disfavors the $\Lambda$CDM model but only at $\sim 1\sigma$, suggesting that in the $w_0w_a$CDM parameterization the evidence for dark energy dynamics is partially caused by the excess smoothing of some Planck CMB anisotropy data. This is similar to what happens in the XCDM$+A_L$ parameterization, \cite{deCruzPerez:2024abc}, to which the $w_0w_a$CDM$+A_L$ parameterization reduces to in the large $z$ limit.

These are interesting results, not yet statistically significant, but certainly worth additional study. Importantly $w(z) = w_0 + w_az/(1+z)$ is a parameterization and not a physically consistent dynamical dark energy model. In the simplest physical dynamical dark energy models dark energy is modelled as a dynamical scalar field $\phi$ with a self-interaction potential energy density $V(\phi)$, \cite{Peebles:1987ek, Ratra:1987rm}. For recent discussions of the DESI 2024 data constraints on dynamical scalar field dark energy models, see \cite{Tada:2024znt, Yin:2024hba, Berghaus:2024kra}. More importantly, of course, is the need for more data, which should soon be forthcoming from DESI.

\acknowledgements
C.-G.P.\ was supported by a National Research Foundation of Korea (NRF) grant funded by the Korea government (MSIT) No.\ RS-2023-00246367.\ J.d.C.P.\ was supported by the Margarita Salas fellowship funded by the European Union (NextGenerationEU).

\bibliography{w0waCDM.bbl}

\end{document}